# A Study on the Open Source Digital Library Software's: Special Reference to DSpace, EPrints and Greenstone

Shahkar Tramboo
Department of Library and Information Science
University of Kashmir
Srinagar

Humma
Department of Library and Information Science
University of Kashmir
Srinagar

S M Shafi
Department of Library and Information Science
University of Kashmir
Srinagar

Sumeer Gul, PhD.
Department of Library and Information Science
University of Kashmir
Srinagar

## ABSTRACT
The richness in knowledge has changed access methods for all stake holders in retrieving key knowledge and relevant information. This paper presents a study of three open source digital library management software used to assimilate and disseminate information to world audience. The methodology followed involves online survey and study of related software documentation and associated technical manuals.

## General Terms
Open source Digital Library Management Software, Information Dissemination.

## Keywords
Open source, Digital Library, Digital Library Management Software, Information Dissemination.

## 1. INTRODUCTION
Open source defines method of software development, that harnesses the power of distributed peer review and transparency of progress **[1]**. This technique helps to provide better quality software's having higher reliability, flexibility with lower cost, and an end to the traditional vendor lock-in. The source code and rights that where normally reserved for copyright holders are now being provided under a free software license that permits developers / users to study, change, improve and at times also to distribute the software **[2]**.

Digital library refers to a collection that constitutes electronic resources, accessible through the World Wide Web. It often contains electronic versions of books, photographs, videos that are owned by a "physical" library **[3]**. Open source digital library software presents a system for the construction and presentation of information collections. It helps in building collections with searching and metadata-bases browsing facilities. Moreover, they are easily maintained and can be augmented and rebuilt automatically. With many Open Source Software (OSS) applications now available for library and information management, Organizations now have novel options for acquiring and implementing systems. The Open Source Software applications for library and Information management that will be discussed in this paper are:

- DSpace
- Greenstone
- EPrints

## 2. DIGITAL LIBRARY MANAGEMENT SYSTEMS
Digital Libraries have greatly evolved during the last few years. They are no longer only the digital counter part of physical libraries (or physical museums, video achieves, etc.) rather they are intricate networked systems capable of supporting communication and collaboration among different, worldwide distributed user communities. Digital Library management system evolved with the inception of Digital Library **[4]**. Digital Library management system provides the appropriate framework both for the production and administration of Digital Library System by incorporating functionality essentially fundamental to Digital Libraries, and also provides provision for integration of additional software that provides more refined and advanced functionality. Digital Library can thus be established by setting up and deploying a Digital Library Management System and then loading or harvesting content. This approach largely simplifies and reduces the effort required to set up a Digital Library that promises a guaranteed better quality of service. These generic systems have started to appear from the second half of 1990's even though implementing the devised DLMS features only to some extent. The major characteristics that distinguish them from each other are the class of functionality offered, the type of object model for information being supported, and the openness of their architecture's. The DLMS (Digital Library Management System) available are commercial as well as open source. But, Open Source DLMS's (Digital Library Management System) are the one that will be studied. Open source digital library management software's provide extensible features to administrators' and allows an organization to showcase their digital achieve to world audience. With full rights of software available under GPL and source code being provided with the software, Organization's can extend the functionality of the software as being required for the particular operation. The DLMS' (Digital Library Management System) studied are:

### 2.1 DSpace
The DSpace is a joint project of the MIT Libraries and HP labs **[5]**. It is a digital asset management system that allows institutions, such as libraries to collect, archive, index, and disseminate the scholarly and intellectual efforts of a community. Written with a combination of technologies by MIT, it is primarily used to capture bibliographic information describing articles, papers, theses, and dissertations. DSpace is adaptable to different community needs. Interoperability between systems is built-in and it adheres to international standards for metadata format. Being an open source technology platform, DSpace can be customized to extend its capabilities. Some of its characteristics as shown in DSpace documentation are as:

a) It is a service model for open access and/or digital archiving for perennial access.

b) Provides a platform to frame an Institutional Repository and the collections are searchable and retrievable by the Web.

c) Helps to make available institution-based scholarly material in digital formats. The collections will be open and interoperable.

The organization of data modal in DSpace is intended to mirror the structure of the organization using the DSpace. Each DSpace site is divided into communities, which can be further divided into sub-communities reflecting the typical university structure of college, department, research centre, or laboratory **[6]**. Communities contain collections, which are groupings of related content. A collection may appear in more than one community. Each collection is composed of items, which are the basic archival elements of the archive. Each item is owned by one collection. Additionally, an item may appear in additional collections; however every item has one and only one owning collection. Items are further subdivided into named bundles of bitstreams. Bitstreams are, as the name suggests, streams of bits, usually ordinary computer files. Bitstreams that are somehow closely related (for example HTML files and images that compose a single HTML document) are organized into bundles.

As specified by **Robert Tansley, Mick Bass, Margret Branschofsky, Grace Carpenter, Greg McClellan, David Stuve (05-Oct-2005)** the bundles most items tend to have included the following:

a) ORIGINAL: The bundle contains the original, deposited bitstreams.

b) THUMBNAILS: Thumbnails of any image bitstreams.

c) TEXT : It includes extracted full-text from bitstreams in ORIGINAL, for indexing.

d) LICENSE: It contains the deposit license that the submitter granted the host organization; putting it differently it specifies the rights that the hosting organizations have.

e) CC_LICENSE: It contains the distribution license, if any (a Creative Commons license) associated with the item. This license specifies what end users can do with the downloaded content.

Each bitstream is associated with one Bitstream Format. Because preservation services are an important aspect of the DSpace service, it is important to capture the specific formats of files that users submit **[7]**. The format of bitstream is a unique and provides a coherent way to sort out a particular file format. The implicit or explicit notion of a bitstream format is to provide means how material in that format can be interpreted. For example, the bitstream interpretation for still images compression encoded in the JPEG standard is defined explicitly in the Standard ISO/IEC10918-1. **[8]**.

In DSpace data modal each item has one qualified Dublin Core metadata record. An item may have other metadata stored in as serialized bitstream, but for every time Dublin Core is used to provide interoperability and ease of discovery. The Dublin Core may be entered by end-users as they submit content, or it might be derived from other metadata as part of an ingest process. The removal of items in DSpace is done in two ways: They may be 'withdrawn', which means they remain in the archive but are completely hidden from view. In this case, if an end-user attempts to access the withdrawn item, they are presented with a 'tombstone,' that indicates the item has been removed. For whatever reason, an item may also be 'expunged' if necessary, in which case all traces of it are removed from the archive **[9]**. The features of DSpace as Digital Management Software are as follows:

a) **Authentication:** DSpace allows contributors to limit access to items in DSpace, at both the collection and the individual item level **[10]**. The mechanism whereby the system securely identifies its users.

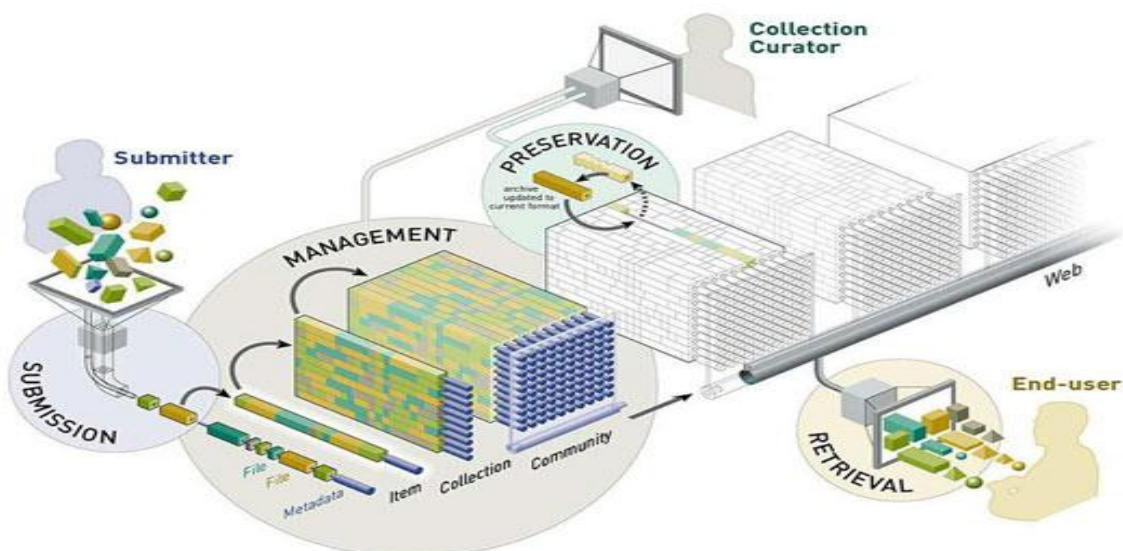

**Fig 1: DSpace Digital repository Modal [6].**

b) **Authorization:** The mechanism by which a DSpace determines what level of access a particular authenticated user should have to secure resources controlled by the system is done by keeping access control policies that allow it to understand what credentials are required (if any) to undertake particular actions upon particular resources **[10]**. Authentication is provided through user passwords, X509 certificates or LDAP. Access controls can be administered by only authorized users. The access controls specify default distribution policy for all items, specify users to submit to collection and specify reviewers, approvers, and metadata editors for a collection's submission process. There are two built-in groups: 'Administrators', who can do anything in a site, and 'Anonymous', which is a list that contains all users. Assigning a policy for an action on an object to anonymous means giving everyone permission to do that action. (For example, most objects in DSpace sites have a policy of 'anonymous' READ.) Permissions must be explicit - lack of an explicit permission results in the default policy of 'deny'. Permissions also do not 'commute'; for example, if an e-person has READ permission on an item, they might not necessarily have READ permission on the bundles and bitstreams in that item. Currently Collections, Communities and Items are discoverable in the browse and search systems regardless of READ authorization.

c) **Non-dynamic HTML document Support:** As mentioned by **Tansley R, et al (2005)** in the documentation, DSpace simply supports uploading and downloading of bitstreams as-is. This mechanism is good for majority of file –formats like PDF, Word Document and so on. As far as HTML documents are concerned they are complicated in the sense they consist of several files and are cross-linked with each other. This has important ramifications when it comes to digital preservation. Web pages also link to or include content from other sites, often imperceptibly to the end-user. Thus, in a few year's time, when someone views the preserved Web site, they will probably find that many links are now broken or refer to other sites than are now out of context. In fact, it may be unclear to an end-user when they are viewing content stored in DSpace and when they are seeing content included from another site, or have navigated to a page that is not stored in DSpace. This problem can manifest when a submitter uploads some HTML content. For example, the HTML document may include an image from an external Web site, or even their local hard drive. When the submitter views the HTML in DSpace, their browser is able to use the reference in the HTML to retrieve the appropriate image, and so to the submitter, the whole HTML document appears to have been deposited correctly. However, later on, when another user tries to view that HTML, their browser might not be able to retrieve the included image since it may have been removed from the external server. Hence the HTML will seem broken.

There is much research going on the issues above. Currently, DSpace bites off a small, tractable chunk of this problem. DSpace can store and provide on-line browsing capability for self-contained, non-dynamic HTML documents. By dynamic content means no CGI script and so on. The links preserved for images, videos etc are preserved as relative links. Any absolute link is stored as is and will continue to link the source as long as it is live, and will eventually change or disappear.

d) **OAI-PMH Support:** The OAI_PMH is a protocol for metadata harvesting. This allows sites to programmatically retrieve or 'harvest' the metadata from several sources, and offer services using that metadata, such as indexing or linking services **[11]**. DSpace exposes the Dublin Core metadata for items that are publicly (anonymously) accessible. Additionally, the collection structure is also exposed via the OAI protocol's 'sets' mechanism. OCLC's open source OAICat framework is used to provide this functionality.
The OAI service can also be configured to make use of any crosswalk plug-in to offer additional metadata formats, such as MODS. Deletion information for withdrawn items is not displayed by DSpace's OAI. DSpace also supports OAI-PMH resumption tokens. Hierarchy to manage contents (i.e. Communities, Collections, and Items).

e) **Object Management:** The process of item ingestion in DSpace is via a web interface or batch item importer. In workflow process for item submission will initiate depending on the configuration of collection. The workflow process may contain one or more steps as per the user need. The collection and communities in DSpace are created via web interface **[6]**.

f) **Import & Export:** Import & Export for Communities, Collections and Items is supported by DSpace. It also includes batch tools to import and export items in a simple directory structure, where the Dublin Core metadata is stored in an XML file. This may be used as the basis for moving content between DSpace and other systems.

g) **Statistics:** Statistics are provided for administrative usage. Statistical reports/summary can be used for performing analysis on repository, providing information like number of items uploaded, searched, number of e-people registered with the system etc **[12]**.

h) **Handle System:** To help in creation of persistent identifier for every item DSpace makes use of Handle system's global resolution feature. DSpace requires a storage and location independent mechanism for creating and maintaining identifiers. DSpace uses the CNRI Handle System for creating these identifiers **[11]**. A Handle server runs as a separate process that receives TCP requests from other Handle servers, and issues resolution requests to a global server or servers if a Handle entered locally does not correspond to some local content **[13]**.

i) **Customization & types of document supported:** DSpace allows customization to accommodate the multidisciplinary and organizational needs of a large institution. Albeit DSpace provides a flexible data object modal. It does not allow construction of very different objects with independent metadata sets due to its database oriented architecture **[14]**. DSpace allows minor intervention of user interface. DSpace collections include audio, video or text depending on the organizational needs. The system can function with many file types, including: PDF, HTML, JPEG, TIFF, MP3, and AVI etc.

j) **Standards Compliance:** The default configuration permits DSpace to store metadata of an item in the Dublin Core Metadata Schema **[14]**. This ensures that data can be exchanged with other standards compliant system, such as MARC21. MARC is an acronym for Machine-Readable Cataloguing.

k) **Optimized Search & Browse:** As per **Bass, M J et al (n.d)**, the system allows end-users to discover content in a number of ways, including:

- By default indexing of basic metadata set qualified DC is provided by DSpace. While as indexing of other metadata sets is provided by Jakarta Lucene search engine. Apache Lucene is written in java and provides high-performance, full-featured text search engine library. It provides technology for any application that requires full-text search, especially cross-platform **(Lucene, 2012)**. Lucene supports fielded search, stemming & stop words removal. By default Browsing in DSpace is by title, author, and date field.

- Via external reference, such as a CNRI Handle. A persistent identifier used for every bitstreams of every item.

## 2.2 Greenstone

Greenstone Digital Library Software is a project from New Zealand that provides a new way of organizing information and making it available over the Internet. Collections of information comprise large numbers of documents (typically several thousand to several million), and a uniform interface is provided to them. Libraries include many collections, individually organized, though bearing a strong family resemblance. A configuration file determines the structure of a collection **[15]**. Existing collections range from newspaper articles to technical documents, from educational journals to oral history, from visual art to videos, from MIDI pop music collections to ethnic folksongs.

A typical digital library built with Greenstone will contain many collections, individually organized. Easily maintained, collections can be augmented and rebuilt automatically **[16]**. There are several ways to find information in most Greenstone collections. For example, you can search for particular words that appear in the text, or within a section of a document. Word search is provided as Greenstone constructs full-text indexes from the document text that is, indexes that enable searching on any words in the full text of the document. Indexes can be searched for particular words, combinations of words, or phrases, and results are ordered according to how relevant they are to the query. In most collections, descriptive data such as author, title, date, keywords, and so on, is associated with each document **[15]**. This information is called metadata. Many document collections also contain full-text indexes of certain kinds of metadata. You can browse documents by title: just click on a book to read it. You can browse documents by subject. Subjects are represented by bookshelves: just click on a bookshelf to look at the books. Where appropriate, documents come complete with a table of contents: you can click on a chapter or subsection to open it, expand the full table of contents, or expand the full document into your browser window **[17]**.

Collections can contain text, pictures, audio and video. Non-textual material is either linked into the textual documents or accompanied by textual descriptions such as figure captions to allow full-text searching and browsing **[18]**. Unicode is used throughout Greenstone. This allows any language to be processed and displayed in a consistent manner. Multilingual collections embody automatic language recognition, and the interface is available in all the major languages **[16]**.

Figure below shows several users, at the top of the diagram, accessing the Greenstone collections. Before going online, these collections undergo the importing and building processes. First, documents, shown at the bottom of the figure, are imported into the XML-compliant Greenstone Archive Format. Then the archive files are built into various searchable indexes and a collection information database that includes the hierarchical structures that support browsing. When this is done, the collection is ready to go online and respond to requests for information.

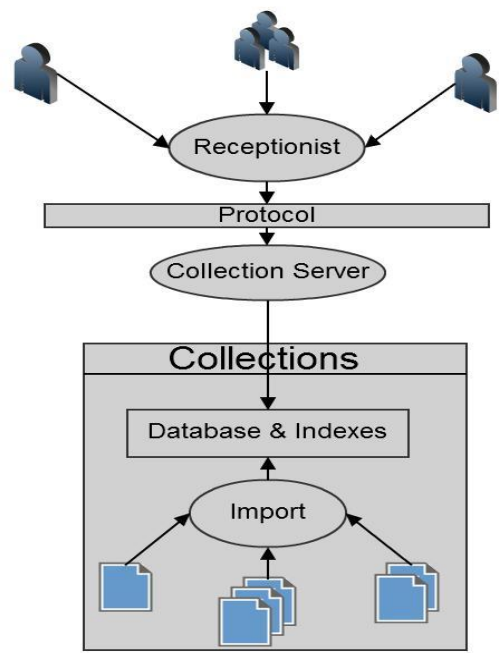

**Fig 2: Data Modal of Greenstone.**

Two components are central to the design of the runtime system: "receptionists" and "collection servers." From a user's point of view, a receptionist is the point of contact with the digital library. It accepts user input, typically in the form of keyboard entry and mouse clicks; analyzes it; and then dispatches a request to the appropriate collection server (or servers) **[19]**.This locates the requested piece of information and returns it to the receptionist for presentation to the user. Collection servers act as an abstract mechanism that handles the content of the collection, while receptionists are responsible for the user interface **[20]**. Receptionists communicate with collection servers using a defined protocol. The implementation of this protocol depends on the computer configuration on which the digital library system is running. The most common case, and the simplest, is when there is one receptionist and one collection server, and both run on the same computer. This is what you get when you install the default Greenstone.

Collections are accessed over the Internet or published, in precisely the same form, on a self-installing Windows CD-ROM **[16]**. Compression is used to compact the text and indexes. A Corba protocol supports distributed collections and graphical query interfaces.

Listed below are some of special features possessed by the Greenstone:

a) **Accessible via web browser:** Collections are accessed through a standard web browser (Netscape or Internet Explorer) and combine easy-to-use browsing with powerful search facilities **[16]**.

b) **Full Text and Field Search:** The user can search the full text of the documents, or choose between indexes built from different parts of the documents **[15]**. For example, some collections have an index of full documents, an index of sections, an index of titles, and an index of authors, each of which can be searched for particular words or phrases. Results can be ranked by relevance or sorted by a metadata element.

c) **Flexible browsing facilities:** The user can browse lists of authors, lists of titles, lists of dates, classification structures, and so on **[15]**. Different collections may offer different browsing facilities and even within a collection, a broad variety of browsing interfaces are available. Browsing and searching interfaces are constructed during the building process, according to collection configuration information **[17]**.

d) **Create access structures automatically:** The Greenstone software creates information collections that are very easy to maintain. All searching and browsing structures are built directly from the documents themselves. No links are inserted by hand, but existing links in originals are maintained **[19]**. This means that if new documents in the same format become available, they can be merged into the collection automatically. Indeed, for some collections this is done by processes that wake up regularly, scout for new material, and rebuild the indexes—all without manual intervention.

e) **Make use of available metadata:** Metadata, which is descriptive information such as author, title, date, keywords, and so on, may be associated with each document, or with individual sections within documents **[20]**. Metadata is used as the raw material for browsing indexes **[21]**. It must be either provided explicitly or derivable automatically from the source documents. The Dublin Core metadata scheme is used for most electronic documents; however, provision is made for other schemes.

f) **Plug-in extends system's capabilities:** In order to accommodate different kinds of source document, the software is organized in such a way that "plug-in" can be written for new document types **[16] [19]**. Plug-in currently exist for plain text, html, Word, PDF, PostScript, E-mail, some proprietary formats, and for recursively traversing directory structures and compressed archives containing such documents.

g) **Customization:** The Greenstone allows customization of presentation of collection that are based on EXtensible Stylesheet Language transformation (XSLT) and other agents that govern the definite functions of Digital library. The architecture of Greenstone purvey:
   a. A back end that provide services to manage documents and collections.
   b. A front end that provides a web based interface for searching and presentation of documents, collections.

h) **Designed for Multi-gigabyte collection:** Collections can contain millions of documents, making the Greenstone system suitable for collections up to several gigabytes **[16]**.

i) **Multilingual Support:** Unicode is used throughout the software, allowing any language to be processed in a consistent manner. To date, collections have been built containing French, Spanish, Maori, Chinese, Arabic and English **[19] [16]**. On-the-fly conversion is used to convert from Unicode to an alphabet supported by the user's web browser.

j) **Collections support multiple formats:** Greenstone collections can contain text, pictures, audio and video clips **[16]**. Most non-textual material is either linked in to the textual documents or accompanied by textual descriptions (such as figure captions) to allow full-text searching and browsing.

k) **Administrative function provided:** An "administrative" function enables specified users to authorize new users to build collections, protect documents so that they can only be accessed by registered users on presentation of a password, examine the composition of all collections, and so on **[16]**. Logs of user activity can record all queries made to every Greenstone collection **[19] [20]**.

l) **Collections can be published on the Internet or on CD-ROM:** The software can be used to serve collections over the World-Wide Web. Greenstone collections can be made available, in precisely the same form, on CD-ROM **[16] [15]**. The user interface is through a standard web browser (Netscape is provided on each disk), and the interaction is identical to accessing the collection on the web—except that response times are more predictable. The CD-ROMs run under all versions of the Windows operating system.

## 2.3 EPrints

EPrints is free software developed by the "University of Southampton, England". EPrints repository collects preserves and disseminates in digital format the research output created by a research community. It enables the community to deposit their preprints; post prints and other scholarly publications using a web interface, and organizes these publications for easy retrieval. It is the world's first, most widely used, and by far the most functional of all the available OA IR software's.

It is created for and specifically focused on OA functionality **[22]**. EPrints is an extensible content management system. It has been extensively configured to accommodate the needs of academics and researchers amid at dissemination and reporting, but it could be easily used for other things such as images, research data, audio archives - anything that can be stored digitally, but you'll have make more changes to the configuration. EPrints is OAI-complaint. It is highly configurable to achieve diverse needs, built on a coding platform that is amendable to rapid development **[23]**.

The real strength of EPrints lies in its ease of use for both end-users and administrators. Submitting documents in EPrints is very straightforward **[24]**. Users are taken through the submission process one step at a time and asked to provide metadata information along with an electronic copy of the document. Users can simply enter metadata such as document type, title, author name, date, etc. via a web form, no knowledge of HTML or XML is required. The metadata fields that appear on the form are selected by the administrator. Administrators can easily customize the metadata form, so that only those fields that are pertinent to a given collection are presented to the end-user. Submissions to achieve can be easily managed by the user, and also editing, updating, and removal of documents is possible after submission (although the administrator can limit these functions). Browsing in EPrints can be done based on any of the metadata fields within a collection, and multiple browsing criteria can be used. For example, in browsing a collection of theses, it would be possible to browse by department and then break down the results by supervisor and year. The browsing categories that are made available to the user are controlled by the administrator. EPrints build and manage OAI-compliant EPrints archives **[25]**. The documents in an EPrints archive can be indexed to allow retrieval by online search engines like Google, which helps to ensure greater access to, and greater dissemination of any items uploaded to the archive. Searching is fairly limited in EPrints. As mentioned earlier, Boolean searching is not supported **[26]**. It is also quite easy to run a search that yields no results. For end users accustomed to modern search engines and databases it might be discouraging to get an unsuccessful search with no suggestions for alternative search strategies.

In EPrints there is no such strict structural division into sections and collections that are still playing an important role, for example, to narrow the search to the repository. The idea is that all records are equivalent and do not form a hierarchy. Nevertheless, the hierarchy is needed to navigate through the repository, since the users can not exactly know the purpose of their search, having only a rough idea about it. The way to generate the navigation of any desired type using the related elements of the metadata fields, i.e. In EPrints the problem is solved using the so-called representations (views). A way to generate the navigation of any desired type using the related elements of the metadata fields, i.e. representation can be carried out on parts of the organization or by the author, or a more complex version, the year of publication, and then by Type, etc. Thus, in the EPrints data model it is possible to ensure flexible support to the hierarchical subject classification (according to silence, classification of the library of congress) and the tree of the subdivisions of organization. Such objects, as element, collection of files, file, are similar to the analogous into DSpace. Element is also the fundamental unit of storage and contains all metadata, allowed for the external use. The hierarchical structure of elements is significantly different. DSpace uses a more rigid system, although it covers most of the needs of the repository. EPrints allows you to create a more complex hierarchy based on different external representations.

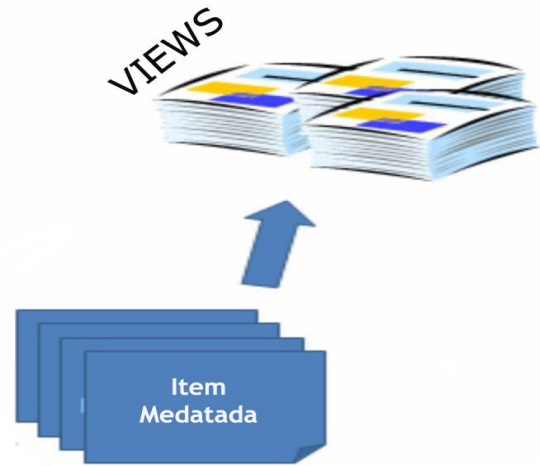

**Fig 3: EPrints Data Modal**

Listed below are some of special features possessed by the EPrints:

a) **Accessibility via web browser:** EPrints provides web based interface that makes it easy to use and administer.
b) **Full Text and Field Search:** Searching is based on metadata not full text based search is supported by EPrints **[27]**. Searching in EPrints allows scanning each of the metadata field types in the database by using simple or advanced search. Any metadata field can be searched with fine granularity by SQL querying the database.
c) **Administrative function provided:** EPrints archive can use any metadata schema as being provided by the administrator. The administrator decides what metadata fields are held about each EPrints item **[28]**. This is specified in three or four stages:
   a. Definition of a maximal set of metadata fields that should be stored (e.g. authors, title, journal, journal volume, etc.).
   b. Definition of different types of EPrints (e.g. refereed journal article, thesis, technical report, unpublished preprint, etc.).
   c. Specification for each type which metadata fields should be stored, and which of those fields are mandatory.
   d. Decide how these metadata fields should be projected into the Open Archives world. (If necessary, interoperability can be switched off, but this is strongly discouraged.)

d) **Open Source Software:** EPrints uses traditional technologies and runs on pure Open Source systems. It uses MySQL, Apache database and web server. MySQL is the world's most popular open source database, recognized for its speed and reliability and Apache has been the most popular web server on the Internet since April of 1996.Eprints is programmed by using the script language "Perl", that is low level but powerful.

e) Three **user roles:** administrator, editor and author.
   a. Administrator role controls all back-end options such as organization of records, web interface appearance and functionality, and all other server-side settings.
   b. Editor role reviews submissions before they are published online and may edit metadata on submissions to maintain consistency or correct errors.
   c. Author role allows submission of documents and management of previously submitted documents.

f) **OAI-PMH Support:** EAS is fully interoperable with OAI (Open Archives Initiative) Protocol for Metadata Harvesting **[29]**. Open Archives protocol allows sites to programmatically retrieve or 'harvest' the metadata from several sources, and offer services using that metadata, such as indexing or linking services. Such a service allows e-prints servers create the potential for a global network of cross-searchable research information, by allowing the contents of servers around the world searched simultaneously by using the OAI (Open Archives Initiative) protocol.

g) **Multilingual Support:** Unicode is used throughout the software, allowing any language to be processed in a consistent manner **[30]**.

h) **File formats supported:** Functions with many file types, including: PDF, HTML, JPEG, TIFF, MP3, and AVI etc. Metadata schema can be tailored to meet the requirements **[27]**.

i) **Statistics:** Statistics are provided for administrative usage .Statistical reports/summary can be used for performing analysis on repository **[30]**.

j) **Customization:** The EPrints data modal consist of user defined metadata. In order to export data in other formats plug-ins can be written. For developers who wish to access the core Digital Library functionality Core API in Perl language is provided **[30]**.

k) **Item preview in EPrints:** Thumbnail preview of documents and images is generated automatically upon file upload **[30]**.

## 3. Based on above discussion a Product Comparison Table for DSpace, EPrints & Greenstone is drafted below

| Feature | DSpace | EPrints | Greenstone |
|---|---|---|---|
| Year of creation | 2002 | 2000 | 1997 |
| License cost | Free | Free | Free |
| Product Type | Software | Software | Software |
| Update cost | Free | Free | Free |
| Resource Identifier | CNRI Handles | No | OAI Identifier |
| OAI-PMH | Yes | Yes | Yes |
| Supported Item Types (Storage *and* rendition) | Can store and manage all types of content | Can store and manage all types of content | Can store and manage all types of content |
| Metadata formats | Dublin Core, Qualified DC, METS | Dublin Core, METS | Dublin Core, Qualified DC, METS, NZGLS (New Zealand Government Locator Service), AGLS (Australian Government Locator Service) |
| User interface functions | End user depositions, Multilingual support. | End user depositions, Multilingual support. | End user deposition, Multilingual support. |
| Thumbnail Preview | Images | Images, Audio, Video | Images, Audio, Video |
| Searching Capabilities | Field Specific, Boolean Logic, Sorting options | Field Specific, Sorting options | Field Specific, Boolean Logic, |
| Browsing options | By Author, Title, Subject and | Browsing can be done using | Browsing can be done |
| | collection | any field. | using any field. |
| Syndication | RSS, ATOM | RSS, ATOM | --- |
| User Authentication | LDAP Authentication, *Shiboleth Authentication* | LDAP Authentication | User Groups |
| Statistical reporting | Count of Full Records | Count of Full records | Count of Full records |
| Software Platforms | Linux or Unix, Solaris, Windows | Linux, Unix, Windows, | Linux, Unix, Windows, Mac-OS |
| Databases | Oracle, PostgreSQL | MySQL, Oracle, PostgreSQL, Cloud. | Its Own |
| Programming Language | Java and JSP | Perl | C++, Perl, Java |
| Web Server | Apache and Tomcat | Apache | Apache/IIS |
| Associated Software | Jave, Apache, PostgreSQL, or Oracle | Linux or Unix, Apache, Perl | Apache, PERL, GNU C++ Compiler, JAVA, GNU Database manager |
| Machine-to-Machine Interoperability. | OAI-MHP, OAI-ORE, SWORD, SWAP | OAI-MHP, OAI-ORE, SWORD, SWAP, RDF | Z39.50, OAI-MHP |
| License | GNU | BSD | GNU |
| Services | Service via 3rd part service providers | Training, Consultancy, Site Visits. | Training. |

## 4. Conclusion

The Digital Library Management softwares (DLMS) present an easy to use, customizable architecture to create online digital libraries. With these institutions/organizations can disseminate their research work, manuscripts, or any other digital media for preservations and world over dissemination of digital items. The software's discussed above present different services and architectures. It is difficult to propose one specific DLMS system as the most suitable for all cases. The study can be used as a reference guide by any organization or institute to decide which one will be ideal for creating and showcasing their digital collection. The choice usually depends on type/format of material, distribution of material, software platform and time frame etc for setting up a Digital Library.